\documentclass[aip,pop,reprint]{revtex4-1}
\usepackage{color}
\usepackage{graphicx}
\usepackage{hyperref}

\newcommand{\omti}{1/L_{T_i}}
\newcommand{\omte}{1/L_{T_e}}
\newcommand{\vthi}{v_{\mathrm{th},i}}
\newcommand{\psit}{\psi_t}
\newcommand{\psita}{\psi_{t,\mathrm{lcfs}}}
\newcommand{\shatl}{\hat{s}_\mathrm{loc}}

\begin{document}


\title{Ion temperature gradient turbulence in helical and axisymmetric RFP plasmas}
\author{I. Predebon}
\affiliation{Consorzio RFX, Corso Stati Uniti 4, 35127 Padova, Italy}
\author{P. Xanthopoulos}
\affiliation{Max-Planck-Institut f\"ur Plasmaphysik, Wendelsteinstra{\ss}e 1, 17491 Greifswald, Germany}

\begin{abstract}
Turbulence induced by the ion temperature gradient (ITG) is investigated in the helical and axisymmetric plasma states of a reversed field pinch device by means of gyrokinetic calculations. The two magnetic configurations are systematically compared, both linearly and nonlinearly, in order to evaluate the impact of the geometry on the instability and its ensuing transport, as well as on the production of zonal flows. Despite its enhanced confinement, the high-current helical state demonstrates a lower ITG stability threshold compared to the axisymmetric state, and ITG turbulence is expected to become an important contributor to the total heat transport.
\end{abstract}

\date{\today}
\maketitle


\section{Introduction}
\label{sec:intro}

A long standing question in the context of magnetic fusion is the impact of 3D shaping on the fundamental properties of the plasma. The two most studied toroidal configurations, stellarators and tokamaks, manifest in general different magnetohydrodynamic (MHD) stability and neoclassical confinement. Comparisons of this sort have already been addressed~\cite{heland}, and the underlying reasons seem to be well understood. Conversely, the difference in the behavior of turbulent transport is a topic which remains to date largely unanswered: while the tokamak line has been extensively investigated in this regard during the last decades, it is only recently that the stellarator community has started taking sophisticated steps in this direction, due to the difficulty in dealing with helical fields~\cite{nunami,xan1}. 

In this paper we exploit the unique flexibility offered by the reversed field pinch (RFP) device, thanks to its ability to produce both helical and axisymmetric plasmas in the course of a single discharge. The RFP is nominally an axisymmetric configuration, which at low plasma current is characterized by a wide spectrum of resonant MHD modes maintaining an overall (quasi)axisymmetry of the plasma. Interestingly though, the RFP plasma at high currents experiences a transition to a helical state, named single helicity~\cite{carraro13}. Depending on the current intensity, the duration of the helical state can be long enough to reach an equilibrium state, which can be described numerically by special codes like VMEC~\cite{vmec}.

Despite the fact that the stability of the ion temperature gradient (ITG) mode~\cite{romanelli,horton} has already been studied in the RFP, so far only axisymmetric equilibrium models have been taken into account~\cite{guo,pred1,sattin,tangri,carmody}. The main conclusion from these geometrically simplified investigations is that the ITG stability threshold in the RFP is larger than in the tokamak, typically by a factor $R/a$ (with $R$ the major radius and $a$ the minor radius of the torus). An explanation comes from a detailed analysis of the parallel dynamics, showing a relevant Landau damping of the mode due to the short field connection length~\cite{guo}. Nonlinear simulations of ITG turbulence have been performed with and without impurities~\cite{pred2}, showing a relatively low ion heat transport and a significant Dimits shift.

In this work we aim at revisiting these findings, considering what the introduction of a helical deformation may cause to ITG turbulence. To do this, we rely on the massively-parallel Eulerian gyrokinetic code GENE~\cite{gene}, applied to the VMEC helical/axisymmetric RFP equilibria with the aid of the code GIST~\cite{gist}. After introducing the MHD equilibria of the RFP in Sec.~\ref{sec:equilibria}, we present the comparison of the respective ITG modes in Sec.~\ref{sec:linear}, to proceed with the ITG turbulence and the behavior of zonal flows in axisymmetric and helical systems in Sec.~\ref{sec:nonlinear}. We conclude with a short discussion in Sec.~\ref{sec:concl}.


\begin{figure}[b]
  \includegraphics{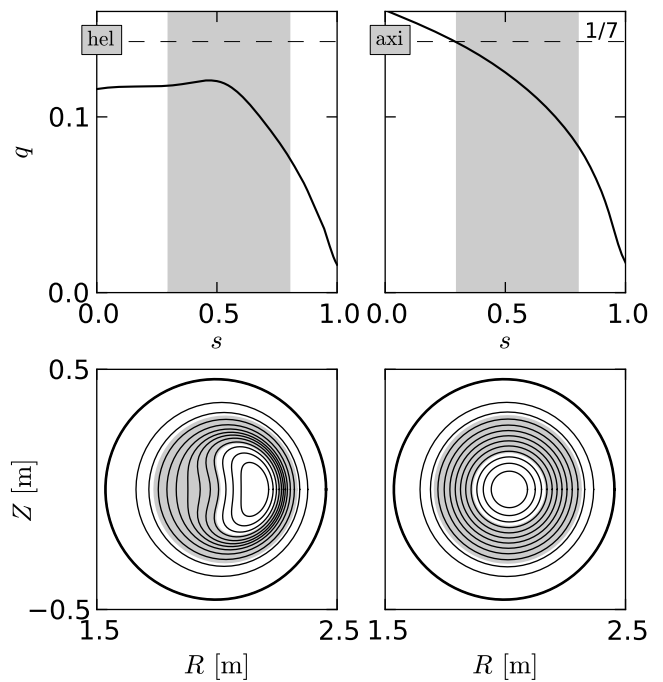}
  \caption{Safety factor profiles (top) and cross-sections at toroidal angle $\phi=0$ (bottom) for the helical (left) and axisymmetric (right) VMEC equilibrium reconstructions. The grey-shaded areas depict the selected radial domain for the gyrokinetic investigation.}
  \label{fig:vmec}
\end{figure}

\begin{figure}[b]
   \includegraphics[width=\columnwidth]{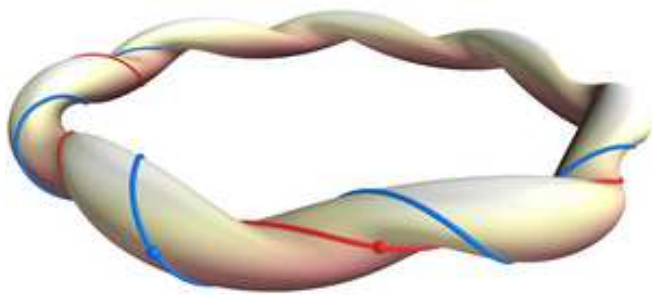}
   \caption{A magnetic flux surface of the helical equilibrium, with the $\alpha=0$ (blue) and the $\alpha=\pi/7$ (red) flux-tubes, where $\alpha=q\theta-\phi$ is the field-label coordinate. The dot on each tube marks the origin in the parallel direction, $z=0$. }
  \label{fig:3d}
\end{figure}

\begin{figure}[b]
  \includegraphics{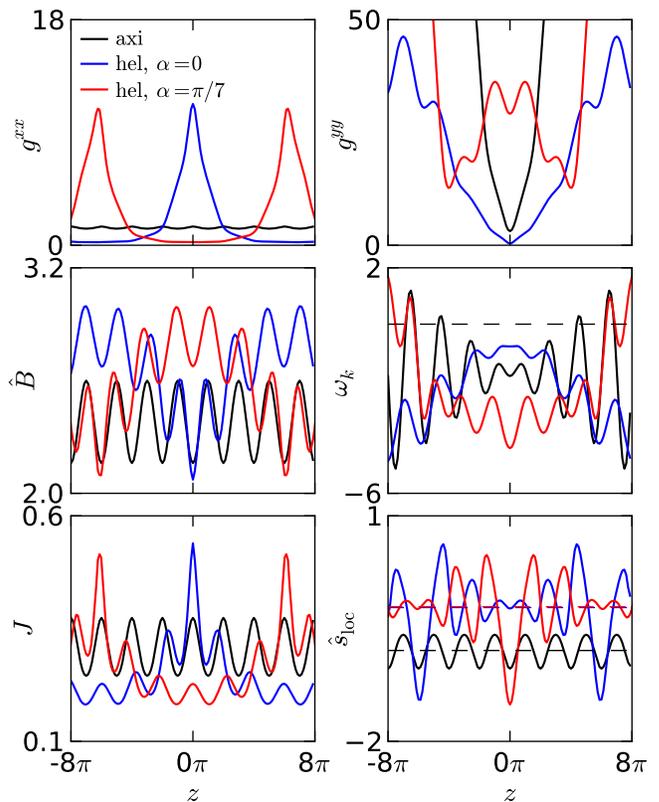}
  \caption{Various geometric coefficients (see text for definitions) versus the parallel coordinate $z$ on the surface $s=0.5$, for the axisymmetric case and for the two stellarator-symmetric tubes, $\alpha=0$ and $\alpha=\pi/7$, for the helical case.}
  \label{fig:gist}
\end{figure}

\begin{figure}[b]
  \includegraphics{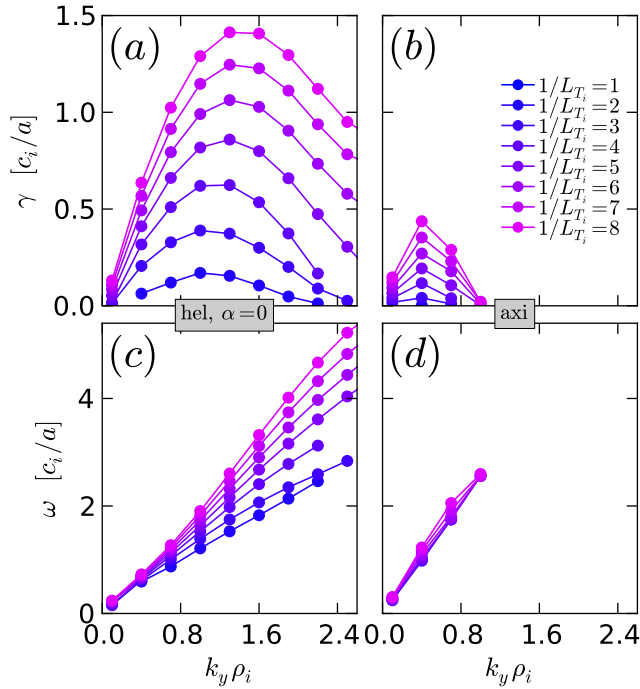}
  \caption{ITG mode growth rate (a)-(b) and real frequency (c)-(d) for the helical $\alpha=0$ tube (left frames) and axisymmetric (right frames) geometry, on the respective $s=0.5$ surface. Here, $c_i=(T_i/m_i)^{1/2}$ is the ion sound speed and $a$ is the minor radius.}
  \label{fig:linear_spec}
\end{figure}

\begin{figure}[b]
  \includegraphics{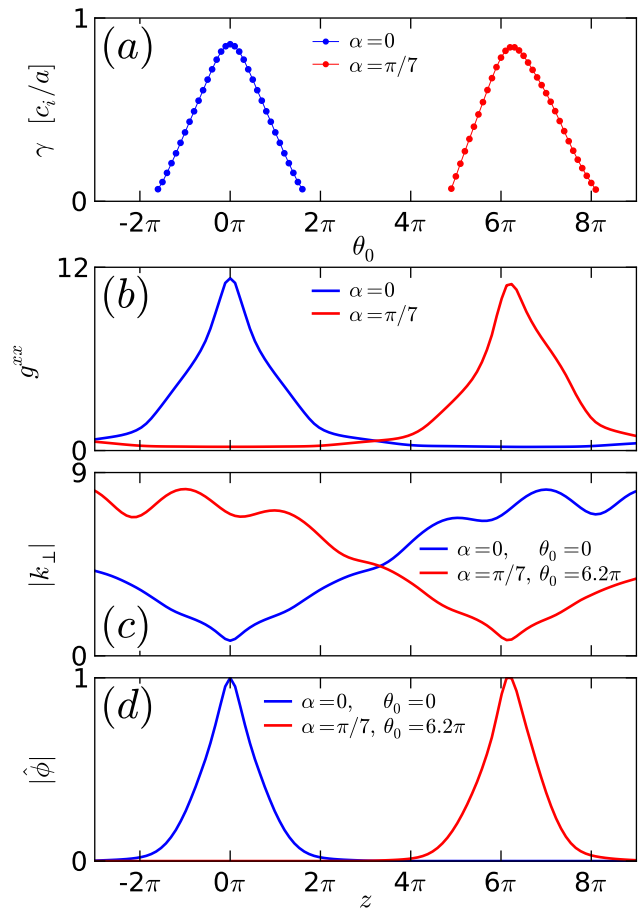}
  \caption{ITG mode growth rate $\gamma$ as a function of the ballooning angle $\theta_0$ for the $\alpha=0$ and $\alpha=\pi/7$ helical flux tubes on the $s=0.5$ surface, for binormal wavenumber $k_y\rho_i=1.2$ and ion-temperature gradient $\omti=5$ (a). The maxima of the $g^{xx}$ component of the metric tensor (b) correspond to the minima of $|k_\perp|$ (with $\theta_0$ such that $\gamma(\theta_0)$ is maximum, see text for definitions) (c), with the structure of the corresponding normalized eigenfunctions in (d).}
  \label{fig:ballooning}
\end{figure}

\begin{figure}[b]
   \includegraphics[width=\columnwidth]{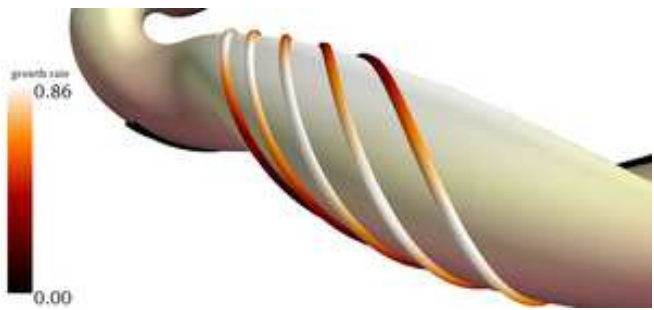}
   \caption{ITG mode growth rate as a function of the ballooning angle for several adjacent flux tubes. The largest growth rates are localized on the external ridge of the helical structure.}
  \label{fig:ballooning3d}
\end{figure}

\begin{figure}[b]
  \includegraphics{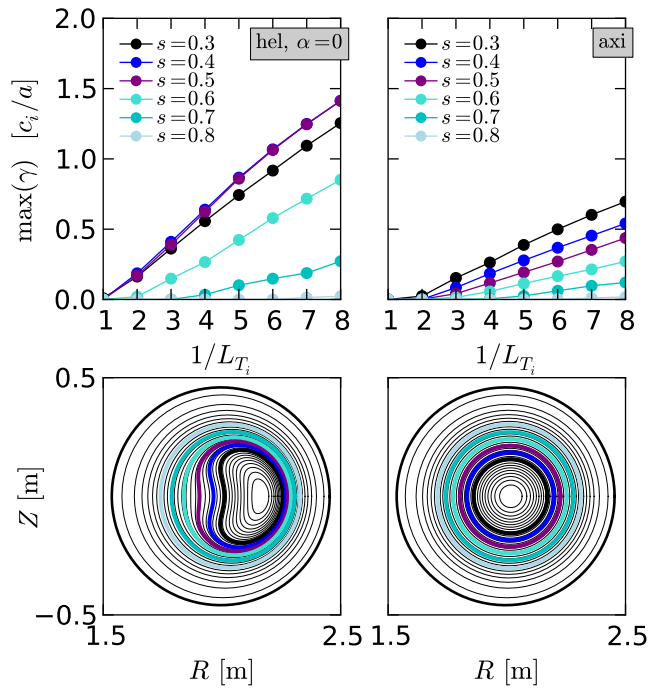}
  \caption{ITG growth rate as a function of the ion temperature gradient across the radial domain $s\in[0.3,0.8]$ for the helical $\alpha=0$ tube (left frames) and axisymmetric (right frames) geometry.}
  \label{fig:linear_s}
\end{figure}

\begin{figure}[b]
  \includegraphics{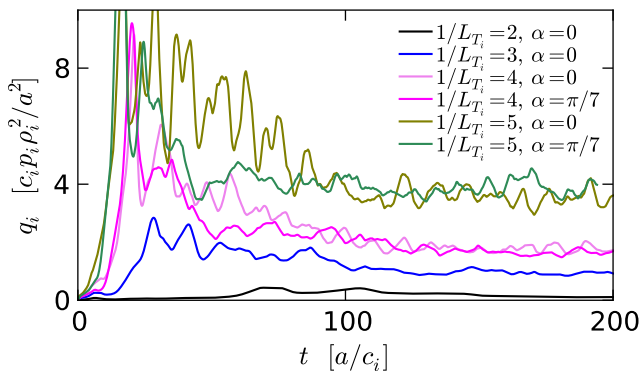}
  \caption{Ion heat flux (normalized to gyro-Bohm units) versus time for the helical geometry, on the $s=0.5$ surface, for different ion temperature gradients, for the $\alpha=0$ and $\alpha=\pi/7$ tubes.}
  \label{fig:qi_t}
\end{figure}

\begin{figure}[b]
  \includegraphics{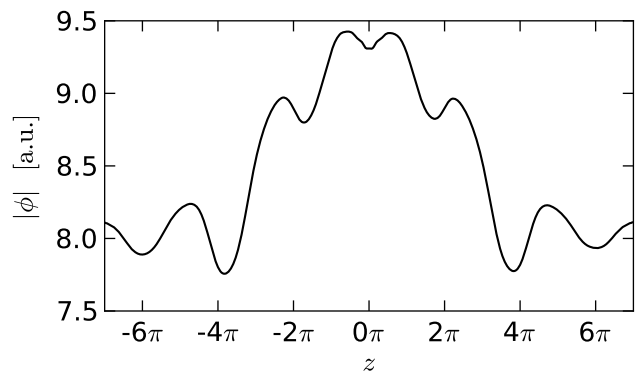}
  \caption{Parallel structure of the absolute value of the time-averaged turbulent potential fluctuations for the helical tube with $\alpha=0$ on the $s=0.5$ surface.}
  \label{fig:phi_z}
\end{figure}

\begin{figure}[b]
  \includegraphics{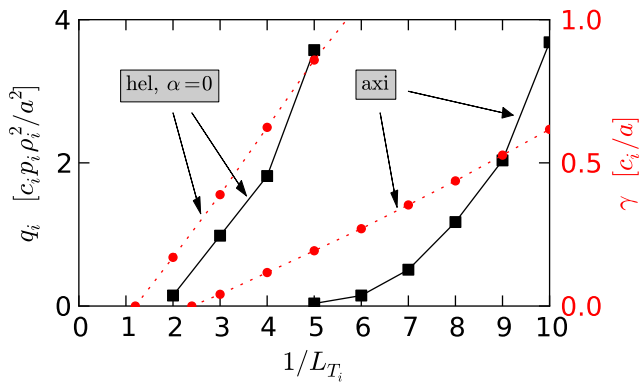}
  \caption{Ion heat flux scaling versus the ion temperature gradient, on the $s=0.5$ surface, for the helical and the axisymmetric tubes. The linear growth rates are also plotted (red), suggesting the usual nonlinear shift, which is particularly pronounced in the axisymmetric case.}
  \label{fig:qi_lti}
\end{figure}

\begin{figure}[b]
  \includegraphics{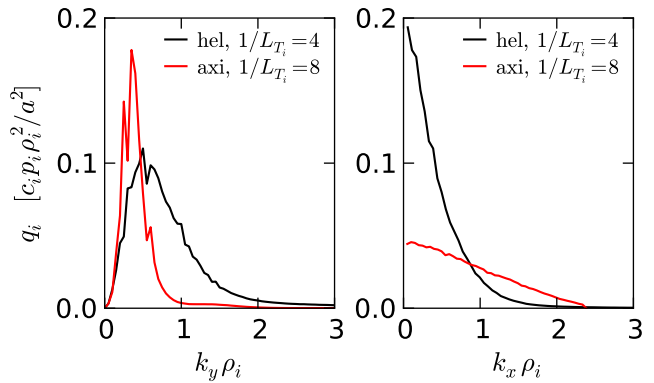}
  \caption{Spectra of the time-averaged ion heat flux in $k_y/k_x$ space, on the $s=0.5$ surface, for the helical $\alpha=0$ and the axisymmetric tubes.}
  \label{fig:nl_spec}
\end{figure}

\begin{figure}[b]
  \includegraphics[width=\columnwidth]{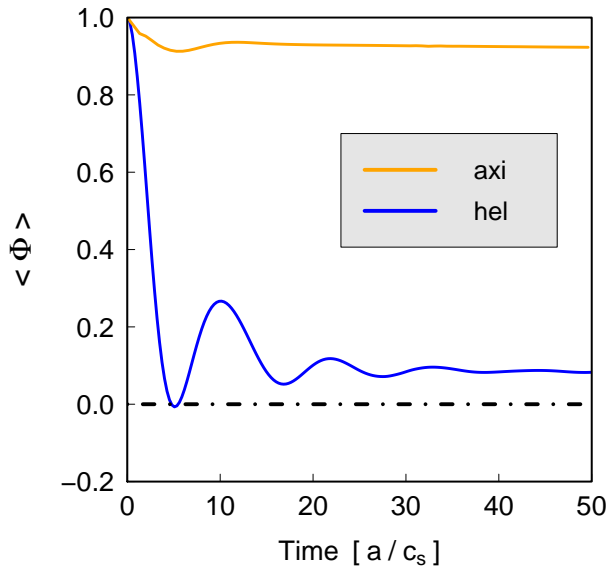}
  \caption{Linear zonal flow response for the helical $\alpha=0$ tube and the axisymmetric one for $k_x\rho_i=0.2\pi$, on the $s=0.5$ surface.}
  \label{fig:zf}
\end{figure}

\begin{figure}[b]
  \includegraphics{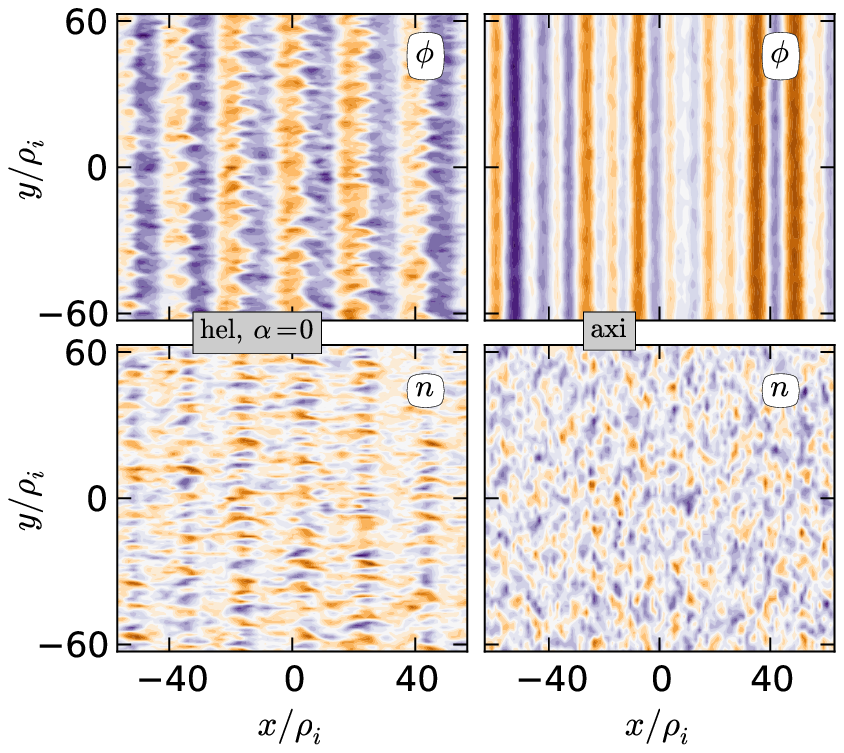}
  \caption{Snapshot of the perturbed electrostatic potential and density, for the helical and axisymmetric tubes.}
  \label{fig:snap}
\end{figure}


\section{Axisymmetric and helical equilibria}
\label{sec:equilibria}

The RFP configuration is characterized by a low safety factor profile, $q\lesssim 0.1$ in the core, which further decreases in the outer region, reaching slightly negative values at the very edge. In case the plasma can be assumed axisymmetric, the connection length of the field turns out to be $L_c= 2\pi R(q^2+\epsilon^2)^{1/2}$, with $\epsilon=r/R$, which is much lower than in a tokamak; in particular $L_c\simeq 2\pi a$ for $r\simeq a$, where the field is purely poloidal.

In RFX-mod, pushing large plasma currents ($\gtrsim$ 1 MA) makes the plasma undergo a transition to a helical state, named single helicity, with a single tearing mode saturating well above the others. Axisymmetric and helical RFP plasmas have rather different features in terms of overall transport properties, MHD dynamics, impurity behaviour~\cite{carraro13}. We mention here only the sharp reduction of magnetic field line stochasticity in the helical states, with the occurrence of transport barriers, at least in the electron heat channel. For this reason, we expect turbulence (at least temperature-gradient driven) to play a potential role in such states.

From the point of view of MHD, the most relevant parameter differentiating the axisymmetric from the single-helicity state is the $q$ profile. Axisymmetric states have a monotonically decreasing $q$ as a function of the radius; for the helical equilibria the $q$ profile has a maximum, which statistically corresponds to the maximum electron temperature gradient~\cite{gobb1}. There has been a considerable effort of the RFX-mod team to describe such equilibria by means of the equilibrium code VMEC, and more recently by means of V3FIT~\cite{terr1,terr2}, which uses VMEC as a solver to determine the equilibrium that best matches the experimental data. In particular, VMEC has been modified to work in the RFP: in a reversed field configuration the toroidal flux $\psit$ cannot be used as a radial coordinate in this configuration, as it is not monotonic. For this reason a new version of the code has been released, which uses the poloidal flux $\psi$ as the radial coordinate.

However, since the magnetic field representation in GIST makes use of the toroidal flux as a radial coordinate, we prefer to slightly modify the $q$ profile in the very edge, keeping it slightly positive. This approximation does not influence the $q$ profile and other equilibrium-related quantities in the radial region where we are going to perform our study, i.e., the mid-radius region where transport barriers emerge. Thus, in the following, the toroidal flux normalized at the last closed magnetic surface, $s=\psit/\psita$, $s\in[0,1]$, will play the role of the radial label.

In Fig.~\ref{fig:vmec} we show two equilibrium reconstructions. On the left, the VMEC reconstruction is made using V3FIT, therefore a minimization procedure is performed on the experimental data. On the right, the reconstruction is based on an axisymmetric $q$ profile~\cite{zanca}, i.e., it does not make use of the $(m,n)=(1,7)$ tearing mode which drives the helical deformation. We prefer to use two different reconstructions at the same time instead of considering different time instants of a discharge, so as to keep most of the plasma parameters equal.

The radial region where we are performing gyrokinetic simulations is $s\in[0.3,0.8]$. The majority of them are on the $s=0.5$ surface, where the (helical) $q$ profile peaks and the magnetic shear is almost vanishing. The GIST code prepares a flux tube domain --with the full geometric description in terms of metric coefficients, curvature operators, Jacobian, magnetic field strength, and parallel gradient-- which the GENE simulation is based on. 

In principle, it is possible to obtain either a Boozer or a PEST coordinate representation of the flux tube. We will use the latter in the following, as that representation is more directly linked to the VMEC $(s,\vartheta,\phi)$ coordinates, $\vartheta$ and $\phi$ being the VMEC poloidal angle and the cylindrical toroidal angle, respectively. In particular, GIST makes use of the straight-field-line poloidal angle $\theta=\vartheta+\lambda(s,\vartheta,\phi)$, with $\lambda$ the stream-function of VMEC. Each point in the flux-tube domain at $s=s_0$ is described by a triplet $(x,y,z)$, with $x=s^{1/2}$ (radial coordinate), $y=(s^{1/2}/q)\,(q\theta-\phi)|_{s=s_0}\equiv(s^{1/2}/q)|_{s=s_0}\,\alpha$ (binormal), $z=\theta$ (parallel).

Due to the peculiar helicity of the helical states, $(m,n)=(1,7)$, the overall magnetic geometry has a seven-fold symmetry. Two stellarator-symmetric~\cite{dewarhudson} flux tubes are used for gyrokinetic studies, one centered in the outboard midplane of the $\phi=0$ section, the other one centered in the outboard midplane at $\phi=-\pi/7$, see Fig.~\ref{fig:3d}. The two tubes have $\alpha=0$ and $\alpha=\pi/7$ respectively. In Fig.~\ref{fig:gist} we show some relevant (normalized) parameters as a function of the parallel flux-tube coordinate $z$~\cite{gist}: the $g^{xx}=|\nabla x|^2$ and $g^{yy}=|\nabla y|^2$ components of the metric tensor, the  normalized magnetic field $\hat{B}$ and its curvature $\omega_k=(\hat{B}\sqrt{g^{xx}}\kappa_\mathrm{norm} + g^{xy}\kappa_\mathrm{geo})/g^{xx}$ (with $\kappa_\mathrm{norm}$ and $\kappa_\mathrm{geo}$ the normal and geodesic components of the field line curvature, respectively; negative $\omega_k$ corresponds to unfavorable curvature causing ITG instability), the Jacobian $J=(\nabla x \times\nabla y \cdot \nabla z)^{-1} $, and the local magnetic shear $\shatl=d(g^{xy}/g^{xx})/dz$, with the dashed lines representing the values of the (global) magnetic shear, given by $\hat{s}=\langle\shatl\rangle_z = 2s\,q'(s)/q$. The even-parity with respect to $z=0$ reflects in the stellarator symmetry of the tubes. The other case shown in the figure is the axisymmetric configuration, where, of course, $\alpha$ does not play any role.


\section{ITG instabilities}
\label{sec:linear}

In RFX-mod, a local estimate of the ion temperature gradient is not available in the plasma core. Ion temperature profiles have been obtained recently from neutral-particle-analyzer data, these profiles being reliable in the edge only, because of the increased neutral absorption probability experienced towards the plasma center~\cite{auriemma}. The Doppler broadening of spectral lines from impurities provides, too, good local estimates of $T_i$ in the outer plasma only, where a ratio $T_i/T_e\sim 2/3$ is commonly evaluated\cite{carraro09}. Therefore, we consider the electron temperature $T_e(s)$ as a reference profile in the core. A detailed study of the dynamics of the electron temperature barriers arising during the helical states can be found in Ref.~\onlinecite{gobb2}. In particular, the $T_e$ barriers turn out to have logarithmic gradients $\omte=-T_e^{-1}\,dT_e/ds^{1/2} \sim 2-3$. Our studies on the ion temperature gradient must cover a larger interval.

In the previous section we have seen how the flux-tube domain is properly defined. The code GENE can now solve the system of gyrokinetic equations to investigate the electrostatic ITG mode. Since we are emphasizing the geometrical aspects of the instability, we consider for simplicity adiabatic electrons. In addition, the density gradient $1/L_n=-n^{-1}\,dn/ds^{1/2}$ is assumed to vanish everywhere. This is a good approximation in the region $s\leq 0.8$, as the experimental density profiles turn out to be essentially flat both in the axisymmetric and in the helical state. High density gradients do exist in the edge, but that domain is outside the scope of this work. Other assumptions are zero collisionality and plasma $\beta$. While the ion collisionality $\nu$ does not introduce relevant differences in the linear results (in RFX-mod experimental conditions, $\nu/(\vthi/a)\sim10^{-3}-10^{-2}$, with $T_i\sim 2/3\; T_e$), assuming vanishing $\beta$ can be more questionable (in RFX-mod, $\beta\sim 1-2\%$). As is known, a finite $\beta$ stabilizes the ITG mode both in the tokamak and the stellarator, see, e.g., Ref.~\onlinecite{ishi} for a recent study on this topic. In axisymmetric RFPs, the $\beta$ suppression turns out to be moderate with respect to the tokamak~\cite{carmody}, with ITG modes being unstable until $\beta\simeq 10\%$. Whether this slow quenching is a feature also of a helical RFP is an open issue, left for a future work.

Let us consider two cases, the helical equilibrium with flux tube $\alpha=0$ and the axisymmetric one, both on the surface $s=0.5$. The other tube for the helical case ($\alpha=\pi/7$) will be considered in a while. The parallel domain is set to cover $z\in[-8\pi,+8\pi]$. In the velocity space, the box size in the $v_\parallel$ direction is $3\,\vthi$ with 32 grid points, while in the $\mu$ direction $9\,T_i/B_0$ with 8 grid points. These conditions are chosen both for the helical and the axisymmetric equilibria, even though shorter domains in the parallel direction are typically required for the axisymmetric equilibria.

Increasing the ion temperature gradient $\omti=-T_i^{-1}\,dT_i/ds^{1/2}$, all the other parameters fixed, the linear spectra in Fig.~\ref{fig:linear_spec} are obtained; the introduction of a helical deformation strongly destabilizes ITG modes. To understand this feature, let us recall that in the ballooning representation~\cite{dewar} perturbed quantities vary as $\delta=\hat\delta(\theta)\exp(iS)$, where $\hat\delta$ is slowly varying in the parallel direction and $S=S(\psi,\alpha)$ describes the rapid variation across the field line. The perpendicular wave vector is $\nabla S=k_\perp=k_\psi\nabla\psi+k_\alpha\nabla\alpha$, the components $k_\psi=\partial_\psi S$ and $k_\alpha=\partial_\alpha S$ being constants. Given $S=k_\alpha(\alpha-q\theta_0)$, with $\theta_0$ the ballooning angle, it follows that $k_\psi=-k_\alpha q'\theta_0$ (or, in normalized units, $k_x=-k_y \hat s\theta_0$, with $\hat{s}$ magnetic shear). Thus $k_\perp^2=k_\alpha^2(g^{\alpha\alpha}-2q'\theta_0 g^{\psi\alpha} + q'^2\theta_0^2 g^{\psi\psi})$. For $\theta_0=0$, which is the case of Fig.~\ref{fig:linear_spec} (this choice will be justified later), the wave vector is reduced to $k_\perp^2=k_\alpha^2 g^{\alpha\alpha}$. As can be seen in Fig.~\ref{fig:gist}, the metric component $g^{\alpha\alpha}$ is much smaller in the helical case than in the axisymmetric case. This implies a reduced finite Larmor radius (FLR) suppression of the ITG mode via the Bessel function $J_0(k_\perp v_\perp/\Omega)$ in the linear gyrokinetic equation. Even if the helical curvature is not so unfavorable as the axisymmetric one --about one half where the mode peaks-- the reduced FLR suppression justifies the higher growth rate instabilities in the helical case. It is useful to remark that the $g^{\psi\psi}$ metric component, which does not enter the linear gyrokinetic equation if $\theta_0=0$, determines the magnitude of $g^{\alpha\alpha}$ via the relation $g^{\alpha\alpha}=(B^2+{g^{\psi\alpha}}^2)/g^{\psi\psi}$ (which is a direct consequence of the Clebsch representation of the magnetic field $B=\nabla\psi\times\nabla\alpha$). Where the ITG mode peaks, $z\in[-\pi,+\pi]$, we have $g^{\psi\alpha}\simeq 0$, so that $g^{\alpha\alpha}\propto 1/g^{\psi\psi}$.

The different weight of the FLR suppression is not the only mechanism acting differently on the two geometries under consideration. As already mentioned in the previous section, the connection length for an axisymmetric RFP is $L_c=2\pi R(q^2+\epsilon^2)^{1/2}$. Such a low connection length is responsible for the mode suppression via Landau damping at low wavenumbers, and for the high ITG stability threshold in axisymmetric geometry~\cite{guo}. In a helical RFP, $L_c$ is not the length of a field line corresponding to a poloidal turn. Contrary to the stellarator case~\cite{plunk}, where $L_c$ is usually considered as the distance between the (stabilizing) spikes of $\shatl$, in the RFP case the local shear has a more oscillatory behaviour (see Fig.~\ref{fig:gist}), so it makes sense to consider for $L_c$ the distance along the field line between two consecutive (destabilizing) peaks of $g^{\psi\psi}$. As such, we typically have $L_{c}^{\mathrm{hel}}\gg L_{c}^{\mathrm{axi}}$, thus $k_{\parallel}^{\mathrm{hel}} \ll k_{\parallel}^{\mathrm{axi}}$. This local property makes the mode suppression via Landau damping less effective in the helical case.

It is known that the (global) magnetic shear $\hat{s}$ influences ITG mode stability. Since the $q$ profile is essentially flat in the helical case for $s\lesssim 0.5$, this provides mode destabilization in the inner region. However, this effect is only marginal with respect to the others mentioned above: forcing axisymmetry with a helical $q$ profile yields a slightly higher growth rate with respect to the axisymmetric $q$, without modifying the wavenumber range where destabilization occurs, $k_y\rho_i\lesssim 1$, see Fig.~\ref{fig:linear_spec}(b).

We can now compare the two helical tubes for linear ITG stability. Starting from the $\gamma(k_y)$ spectrum of the $\alpha=0$ tube, we can fix the binormal wavenumber such that the growth rate is maximum, $k_y \rho_i \sim 1.2$ in our case. Varying $\theta_0$ (or, equivalently, $k_x$), it is confirmed that the maximum $\gamma$ corresponds to $\theta_0=0$ (i.e., $k_x=0$), see Fig.~\ref{fig:ballooning}-a. For the $\alpha=\pi/7$ tube, $\gamma$ is maximum for a ballooning angle $\theta_0\sim 6.2\pi$, which corresponds to the maximum of $g^{\psi\psi}$ along the field line, Fig.~\ref{fig:ballooning}-a/b. Interestingly, the highest growth rate is approximately the same in the two flux tubes. The dependence of $k_\perp$ on $z$ in the two flux tubes is shown in Fig.~\ref{fig:ballooning}-c, where the value of $\theta_0$ (or, equivalently, of $k_x$) is set such that $\gamma(\theta_0)$ is maximum. For the two tubes the lowest $k_\perp$ is approximately the same. For both tubes of the helical configuration a smaller FLR suppression is observed compared to the axisymmetric case. For these parameters, the structure of the normalized electrostatic potential as a function of the parallel coordinate $z$ is shown in frame (d).

The ITG mode growth rate as a function of the ballooning angle $\theta_0$ is shown for several neighboring flux tubes in Fig.~\ref{fig:ballooning3d}. It is evident that the modes have a ``helical ballooning'' structure, being large in correspondence to the peaks of $|\nabla\psi|$, i.e., along the external ridge of the helical structure. Heuristically, the compression of the magnetic surfaces enhances the local temperature gradients, with a consequent growing instability. This is a common feature to any helical plasma configuration, not only the RFP.

The last piece of the linear analysis on ITG stability refers to the radial dependence of the growth rates, Fig.~\ref{fig:linear_s}. We perform this study for $s\in[0.3,0.8]$, this interval fully including the region where the transport barrier usually arises. Both geometries show a larger growth rate in the core than in the edge. While in the axisymmetric case the growth rate monotonically increases as $s\to 0$, in the helical case it is larger for $s\sim 0.4-0.5$, i.e., in the proximity of the maximum of $q$. The radial domain of largest growth rate is characterized, in both geometries, by an overall increased $g^{\psi\psi}$, thus by a decreased $g^{\alpha\alpha}$ and $|k_\perp|$, and by a decreased global magnetic shear $|\hat s|$ (see Fig.~4 of Ref.~\onlinecite{pred1} for an analogous study with GS2). Again, with respect to the axisymmetric case, the helical one clearly yields higher growth rates in the core, where the two sets of geometric coefficients largely deviate. Conversely, at $s\gtrsim 0.7$ the two configurations become equivalent in terms of ITG stability. The footprint of the helical core is rapidly vanishing in the outer region: the peaks in $g^{\psi\psi}$ are no longer present and the curvature starts to have a tokamak-like oscillating behaviour. ITG modes are strongly stabilized by the (negative) magnetic shear, which is rapidly increasing in absolute value towards the edge.


\section{ITG turbulence}
\label{sec:nonlinear}

Having completed the analysis of the ITG mode stability for the helical and axisymmetric RFPs, we turn to the effect of the geometry on the turbulence. It is well known that, apart from the strength of the instability itself, ITG turbulence is largely regulated by the zonal flows. Therefore, again focusing on the different geometric features, in this section we will also provide results on the linear zonal flow response for each configuration under study.

We provide some details concerning the setup of the nonlinear simulations. For the $\alpha=0$ tube on the $s=0.5$ surface (where $q=0.12$ and $\hat s=-0.22$), the simulations are made with the following discretization in the 5-dimensional space: $n_x \times n_y \times n_z \times n_{v_\parallel} \times n_\mu = 128 \times 64 \times 512 \times 32 \times 8$, with $\Delta k_y\rho_i=0.05$, $\Delta k_x\rho_i=0.055$, and $z\in[-8\pi,+8\pi]$. For the $\alpha=\pi/7$ tube, the $z$ domain must be larger, so as to fully include the peaks in $g^{xx}$, $z\in[-10\pi,+10\pi]$, see Fig.~\ref{fig:ballooning}. Accordingly, in order to allow for a full $k_x$ spectrum resolution (not centered at $k_x=0$) a larger number of $x$ grid points is used, namely $n_x=256$, with $\Delta k_x\rho_i=0.098$. Convergence tests have shown that this phase space resolution is adequate. As one can see, the most evident difference with a ``standard'' (axisymmetric) discretization for ITG turbulence simulations is the large box size in the parallel direction: this choice is of central importance, so as to capture all the geometric details, especially for the $\alpha=\pi/7$ tube. Due to the length of the $z$ domain and to the vanishing collisionality, numerical dissipations providing damping in the parallel direction and mimicking diffusion in velocity space have been included by means of hyper-diffusion terms in the $(z,v_\parallel)$ subspace.

In the axisymmetric case the numerical setup is less demanding. For the box size on the $s=0.5$ surface (where $q=0.125$ and $\hat s=-0.79$) we set $n_x \times n_y \times n_z \times n_{v_\parallel} \times n_\mu = 96 \times 64 \times 32 \times 32 \times 8$, with $\Delta k_y\rho_i=0.05=\Delta k_x\rho_i$, and $z\in[-\pi,+\pi]$. Of course the box size, both in the helical and in the axisymmetric case, may differ from surface to surface, depending on the value of the magnetic shear $\hat s$.

With this simulation setup the time history of the ion heat flux $q_i$ for the helical geometry is shown in Fig.~\ref{fig:qi_t} for several ion temperature gradients. For two of them ($\omti=4$ and 5) we show the time trace for both the $\alpha=0$ and the $\alpha=\pi/7$ tube. For the two tubes the same level of saturated turbulence is observed. As a consequence, the two tubes, as in the linear case, can be considered equivalent in terms of physical results, the ion heat flux in this case. However we remark that all the physical fluctuating quantities, e.g. the electrostatic potential $\phi$, the density $n$, the parallel and perpendicular temperature $T_\parallel$ and $T_\perp$, exhibit a clear modulation in the $z$ direction. For instance, the electrostatic potential is peaked where $g^{xx}$ peaks, see Fig.~\ref{fig:phi_z}.

In this light, from now on we can restrict our study to the $\alpha=0$ flux tube. Performing simulations for several ion temperature gradients allows us to compare linear and nonlinear trends, and in particular the temperature gradient threshold in the two cases. In Fig.~\ref{fig:qi_lti} we focus again on the half-flux surface $s=0.5$. While the linear growth rate (dotted lines) already suggests a sharp distinction between the two geometries, in a nonlinear environment their difference is even more pronounced (solid lines). In both cases a Dimits shift\cite{dimits} exists, and the axisymmetric configuration requires ion temperature gradients which must be $2-3$ times larger than the linear threshold for an effective ITG turbulence to take place. A less pronounced upshift occurs in the helical case. As already mentioned, these gradients in RFX-mod can be compared, as a reference, to the electron temperature profiles only, which are characterized by $\omte\sim 2-3$ in the core of the helical geometry, and lower in the axisymmetric one.

Our conclusion so far is that, for an axisymmetric RFP, ITG turbulence is not likely to be a concern from the point of view of particle/heat transport, at least away from the edge. High current helical states, on the contrary, are more prone to ITG instabilities and present a high level of ion heat flux. Besides the difference in the linear spectrum, discussed in the previous section, in Fig.~\ref{fig:nl_spec} we show the wavenumber spectra of the ion heat flux in the $k_y/k_x$ space, averaged over the remaining coordinates and time. The helical and axisymmetric $\omti$ values are chosen to provide approximately the same ion heat flux, $q_i\sim 1-2$, see Fig.~\ref{fig:qi_lti}. In analogy with the linear result, the helical $k_y$ spectrum peaks at smaller scales than the axisymmetric one, a feature which, based on quasi-linear estimates, is however not enough to compensate for the much larger growth rates.

We turn our attention to the effect of zonal flows (ZF)~\cite{zf}, whose role in controlling ITG turbulence in tokamaks is already established. In addition, in the case of non-axisymmetric configurations, it has been recently shown~\cite{xan2}, that although the nature of ZF can be entirely different compared to tokamaks, as well as among different stellarator designs, their impact is still measurable and beneficial. In Fig.~\ref{fig:zf}, we calculate the linear ZF response for the RFX configuration having selected the normalized radial wavenumber equal to $k_x \rho_i=0.2\pi$. Interestingly, oscillations similar to the ones seen in the other stellarators are also here observed, which are attributed to the radial drift of locally trapped particles, and the residual level takes relatively small values. In this situation, it is suggested that the regulation of ITG turbulence takes place during the dynamical state of ZF evolution and is mildly, if at all, affected by the residual level. In the same figure, we show for comparison the linear ZF response for the axisymmetric RFP surface. Due to the small safety factor and to the toroidal symmetry of the system, the residual turns out to be unusually large. Its value is in agreement with the Rosenbluth-Hinton estimate $\mathcal{K}_\mathrm{RH}=1/(1+1.6\,q^2/\epsilon^{1/2})$ for the undamped zonal flow~\cite{rh,sw}, which for the case shown in the figure yields $\mathcal{K}_\mathrm{RH}|_{s=0.5}\simeq 0.92$.

Coming back to the nonlinear simulations, we conclude with the snapshots of the electrostatic potential and of the density fluctuations in the $x/y$ space in Fig.~\ref{fig:snap}, for the two geometries under consideration (the respective $\omti$ gradients are those of Fig.~\ref{fig:nl_spec}). The colour contours confirm the enhanced generation of the zonal flows in the axisymmetric configuration.


\section{Conclusions}
\label{sec:concl}

While a helical core has generally beneficial consequences on the RFP plasma performance, we have shown that it imposes an unfavorable effect in terms of ITG stability and turbulent transport. Focusing on the role of geometry, thus simplifying the physical problem (we have assumed adiabatic electron response, vanishing plasma $\beta$ and collisionality, flat density profile), it turns out that ITG modes are localized in the proximity of the peaks of $|\nabla\psi|$, i.e. where the magnetic surface compression is higher. Here the local temperature gradients become larger, with a consequent growing instability. From the point of view of turbulence, the maxima of $|\nabla\psi|$ reflect in local maxima of turbulent fluctuations, though turbulence turns out to be quite uniformly spread along the flux tubes. With respect to the linear threshold, a moderate nonlinear upshift occurs, with the ion heat flux rapidly increasing with the ion temperature gradient and a consequent high stiffness of the temperature profile. On the contrary, for an axisymmetric RFP, already more linearly stable to ITG modes, the nonlinear shift is much larger. The different behaviour is confirmed by the very high zonal flow residual level in the axisymmetric case, while helical states exhibit a lower zonal flow residual and a stellarator-like oscillating behaviour of the zonal potential. ITG turbulence is consequently expected to be a scarce contributor to particle transport in an axisymmetric RFP, at least without impurities. The inclusion of more realistic parameters of the plasma and a comparison with experiment are the steps to face henceforth.


\acknowledgments

The authors thank D.~Terranova for providing the helical V3FIT equilibrium. The nonlinear simulations have been performed on the HELIOS supercomputer, Japan.



\end{document}